\documentclass[pra,aps,twocolumn,superscriptaddress,showpacs]{revtex4-1}
\usepackage{braket}
\usepackage{soul}
\usepackage{hyperref}
\usepackage{dsfont}
\usepackage{amssymb}
\usepackage{amsmath}
\usepackage{amsfonts}
\usepackage{mathtools}
\usepackage{lipsum}
\usepackage{graphicx}
\usepackage{dcolumn}
\usepackage[dvipsnames]{xcolor}
\usepackage{bm}
\usepackage{multirow}
\usepackage[normalem]{ulem}
\graphicspath{{fig/}}
\newcommand{\affilITMO}{School of Physics and Engineering, ITMO University, St. Petersburg 197101,  Russia}
\usepackage{physics}
\usepackage{appendix}

\definecolor{linkcolor}{HTML}{0176ba}
\definecolor{urlcolor}{HTML}{0176ba} 
\definecolor{citecolor}{HTML}{900020}
\hypersetup{pdfstartview=FitH,  linkcolor=linkcolor,urlcolor=urlcolor, citecolor=citecolor, colorlinks=true}

\begin{document}

\title[]{Strongly subradiant states in planar atomic arrays}

\author{Ilya Volkov}
 \affiliation{\affilITMO}

\author{Nikita Ustimenko}%
\affiliation{\affilITMO 
}%

\author{Danil Kornovan}%
\affiliation{\affilITMO 
}%

\author{Roman Savelev}
\affiliation{\affilITMO 
}%
\author{Alexandra  Sheremet}
\affiliation{\affilITMO
}%

\author{Mihail Petrov}
  \email{m.petrov@metalab.ifmo.ru}
\affiliation{\affilITMO
}%

\begin{abstract}
The optically trapped ensembles of atoms provide a versatile platform for storing and coherent manipulation of quantum information. However, efficient realization of quantum information processing requires long-lived quantum states protected from the decoherence e.g. via spontaneous emission. Here, we theoretically study collective dipolar oscillations in finite planar arrays of quantum emitters in free space and analyze  mechanisms that govern the emergence of strongly subradiant collective states. We demonstrate that the external coupling between the collective states associated with the symmetry of the array and with the quasi-flat dispersion of the corresponding infinite lattice plays a crucial role in the boost of their radiative lifetime. We show that  among different regular arrangements of the atoms the square atomic arrays support eigenstates with minimal radiative losses that scale with the total number of atoms $N_{tot}$ as $\propto N_{tot}^{-5}$. \\

\end{abstract}


\maketitle
\section{Introduction}

In the recent years, a significant progress has been achieved in the development of artificial quantum interfaces based on arrays of {cold atoms}~\cite{Barredo2016Nov,barredo_synthetic_2018,Gross2017,rui_subradiant_2020}. Such ensembles of qubits being ordered advanced optical manipulation methods in the free space \cite{rui_subradiant_2020,barredo_synthetic_2018} or in the vicinity of nanophotonic structures \cite{Corzo2016,Prasad2020,liedl_collective_2023} have already demonstrated potential in storing, processing and transmitting quantum information \cite{Corzo2019}. However, implementation of these technological achievements in practical devices requires solving a number of fundamental problems such as {storing} and protecting quantum information from decoherence. 

One of the possible ways to overcome these problems is to suppress spontaneous emission of quantum emitters, which can be achieved in the {\it subradiant} collective states in the ensembles of entangled qubits. Such states have been largely explored recently in one-dimensional  atomic chains  coupled to a waveguide \cite{Sheremet2023Mar, Kornovan2019Dec,pennetta_observation_2022}, or  atomic clouds in free space \cite{guerin_subradiance_2016}. However, the advantages of two-dimensional arrays  such as methods of precise optical positioning with high filling factor \cite{Barredo2016Nov}, convenience of transmittance and reflection spectrum measurements \cite{Facchinetti2018Feb, Rui2020Jul}, possible convenient application of Rydberg blockade \cite{Saffman2010Aug}, and many others, resulted in outstanding progress in  optical manipulation over qubit states \cite{srakaew_subwavelength_2023}. Moreover, there have been a number of novel theoretical approaches to photons manipulation with atomic lattices \cite{ballantine_parity-time_2021, ballantine_cooperative_2021, nefedkin_nonreciprocal_2023, pedersen_quantum_2023}.  

\begin{figure}[t]
    \centering
    \includegraphics[width=1\columnwidth]{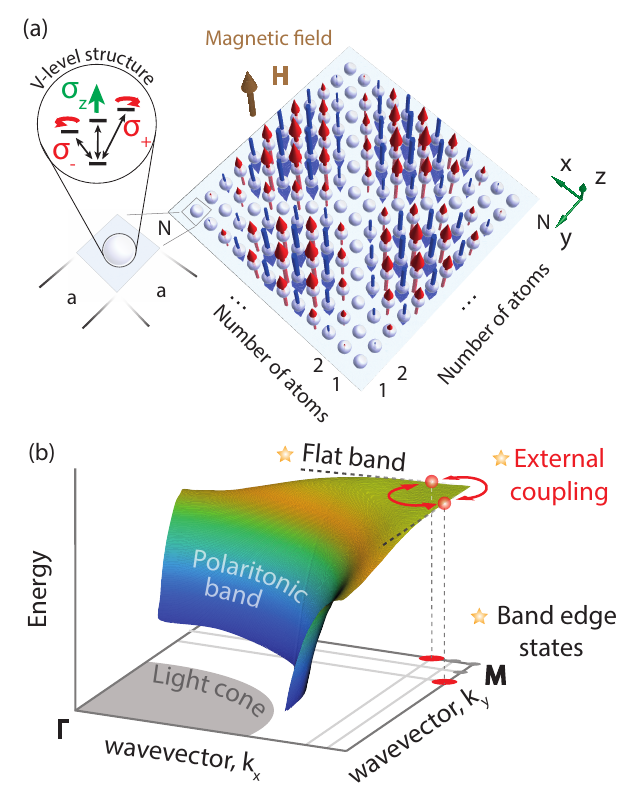}
    \caption{
    (a) A two-dimensional finite square regular lattice of $N\times N$ with $\sigma_z$ transition and separated by distance $a$. The distribution of the highly non-radiative mode of $B_2$ symmetry is shown with arrows.  (b) The polaritonic band diagram of the infinite dipolar lattice. The main mechanisms responsible for the formation of non-radiative states are shown.  }
    \label{fig:fig1_scheme}
\end{figure}

The studies of the long-living states in atomic lattices  were in the center of several works already showing that  in the square two-dimensional (2D) lattices \cite{Facchinetti2016Dec} the eigenstates with out-of-plane dipole moment orientation and in-phase coherence (near the $\Gamma$-point of the Brillouin zone (BZ), see Fig. \ref{fig:fig1_scheme}) was shown to have a lifetime proportional to $N^2$, where $N$ is the number of atoms along one of the directions. Such states fall in the type of bound states in continuum  extensively studied recently in photonics~\cite{Hsu2016,koshelev_bound_2021}, and have been further a subject for investigation in the number of papers \cite{Ballantine2020Apr,de_paz_bound_2023,gladyshev_bound_2022} where their appearance in lattices of different symmetries was also discussed \cite{Sadrieva2019}.

In this work, we consider planar finite arrays of two-level atoms arranged in different regular structures with the focus on the square atomic arrays  schematically shown in Fig.~\ref{fig:fig1_scheme}(a). We identify the main factors affecting the radiative losses suppression in such structures, including interference of the out-of-phase oscillating neighbor dipoles in the far zone, external coupling of the states associated with the symmetry of the structure, and ``accidental'' external coupling that emerges due to quasi-flat polaritonic band dispersion of the corresponding infinite lattice, see Fig.~\ref{fig:fig1_scheme}(b). We establish the conditions under which these mechanisms are present and demonstrate the interplay between them on the example of the square array. We show that the radiative decay rate of the most subradiant states in square arrays decrease as fast as $\propto N_{tot}^{-3}$ or $\propto N_{tot}^{-5}$ ($N_{tot}=N^2$ is the total number of atoms in the array), which leads to very long lifetimes of the eigenstates even in compact arrays. Moreover, there exist optimal geometrical parameters that provide resonant increase of the lifetime by a few orders. We also demonstrate the possibility of selective excitation of the most subradiant states by vector Bessel beams with a suitably chosen orbital angular momentum.

\section{General formalism}

Let us consider an array  of identical two-level atoms in free space with a transition frequency $\omega_0$ (transition wavelength $\lambda_0$) and spontaneous decay rate $\Gamma_0={\omega_0^3 |d|^2}/({3\pi\hbar\varepsilon_0 c^3})$, where  $\varepsilon_0$ is vacuum permeability and $d$ is the amplitude of the transition dipole moment. We assume that the atoms are separated by the distance more than $\sim 0.1\lambda_0$, $\lambda_0 = 2\pi c/\omega_0$, thus neglecting the Casimir interaction between atoms in the ground state due to its very fast spatial decay $\sim 1/r^6$~\cite{Buhmann2015Mar}. Consequently, the coupling between the atoms is determined by the dipole-dipole interaction, governed by the electromagnetic free space Green's tensor $\mathbf{G}(\mathbf{R},\omega)$ \cite{Novotny2012Sep}.

To describe such interaction, we introduce dipole moment operator of the $i$-th atom as $\hat{\mathbf{d}}_i = \mathbf{d}\hat{\sigma}_i + \mathbf{d}^*\hat{\sigma}_i^{\dag} $, where $\hat{\sigma}_i = |g_i\rangle  \langle e_i|$, $\hat{\sigma}_i^{\dag} = |e_i\rangle  \langle g_i|$ are the atomic lowering and rising operators describing the transition from the excited state $|e_i\rangle$ to the ground $|g_i\rangle$ state of the $i$-th atom and vice versa, respectively. 
The interaction Hamiltonian of the system $\hat{\mathcal{H}}_{\text{int}}$, describing both re-scaterring processes governed by dipole-dipole interaction between $N_{tot}$ atoms and spontaneous emission into the free space, can be obtained as the total interaction energy between dipole moments and electric field: $\hat{\mathcal{H}}_{\text{eff}}  = - \sum_{i=1}^{N_{tot}} \hat{\mathbf{d}}_i \hat{\mathbf{E}} (\mathbf{r}_i)$, where $\hat{\mathbf{E}} (\mathbf{r}_i)$ is the electric field operator at the position $\mathbf{r}_i$. Exploiting classical principle of electric field superposition and the connection between the electric field and Green's tensor \cite{Belov2005Aug, Markel2007Feb, AsenjoPRA2017}, one can directly obtain the effective Hamiltonian \cite{reitz_cooperative_2021,ruostekoski_cooperative_2023}:
\begin{multline}     
\hat{\mathcal{H}}_{\text{eff}}  = \hbar \Bigl( \omega_0 - \frac{i}{2}\Gamma_0 \Bigr) \sum\limits_{i=1}^{N_{tot}} \hat{\sigma}_i^{\dag} \hat{\sigma}_i -\\- \sum\limits_{i\neq j} \frac{3\pi \hbar\Gamma_0}{k_0} \mathbf{e_d^*} \cdot \mathbf{G}(\mathbf{R}_{ij})\cdot \mathbf{e_d} \ \hat{\sigma}_i^{\dag} \hat{\sigma}_j,
\label{Hamiltonian}
\end{multline} 
where $\mathbf{e_d} = \mathbf{d}/d$ is the unit vector of the atomic transition dipole moment. In Eq.\eqref{Hamiltonian}, the first term describes the coupling of atoms to free space, while the second one describes the atom-atom coupling. The above equation is obtained within the Markovian approximation by neglecting the frequency dependence of the Green's tensor, $\mathbf{G}(\mathbf{R},\omega) \approx \mathbf{G}(\mathbf{R},\omega_0)$, which is justified by a very narrow resonance of a single atom $\Gamma_0 \ll \omega_0$.

The eigenvalues and eigenstates of the Hamiltonian~\eqref{Hamiltonian} define the complex eigenfrequencies $\omega_j - i \Gamma_j/2$ and eigenfunctions $\Psi^{(j)}$ of the collective dipolar excitations, respectively. Real and imaginary parts of the the eigenfrequency correspond to the frequency and radiative decay rate of the $j$-th collective eigenstate. For convenience throughout the paper we define the frequency with respect to that of a single atom, $\Delta \omega_j = \omega_j - \omega_0$.

Normally, in the absence of the external magnetic field, atoms are spherically symmetric, and the simplest model of such an atom is the one with triply degenerate excited state ($S \leftrightarrow P$ transition). 
In the presence of the $z$-oriented magnetic field [normal to the $x-y$ plane of the array, see Fig.~\ref{fig:fig1_scheme} (a)] the states of an atom can be spectrally separated into three states depending on the polarization of the transition dipole moment, $\sigma_{-}$, $\sigma_{+}$, or $\sigma_z$. Consequently, the collective eigenstates of the planar arrays of such atoms can be considered separately for different polarizations of the atomic transitions.

\section{Square atomic array}

It is known that the radiative losses are mostly suppressed in symmetric structures, which possess fewer radiation channels than non-symmetric ones. In this section, we  focus on one of the most simple and high-symmetry case of 2D atomic array $N\times N$ arranged is a square lattice with a period $a$. Such illustrative example allows us to  demonstrate the competition between different mechanisms of the radiative loss suppression.

\subsection{Classification of the collective eigenstates}

\begin{figure}[t!]
    \centering
    \includegraphics[width=1\columnwidth]{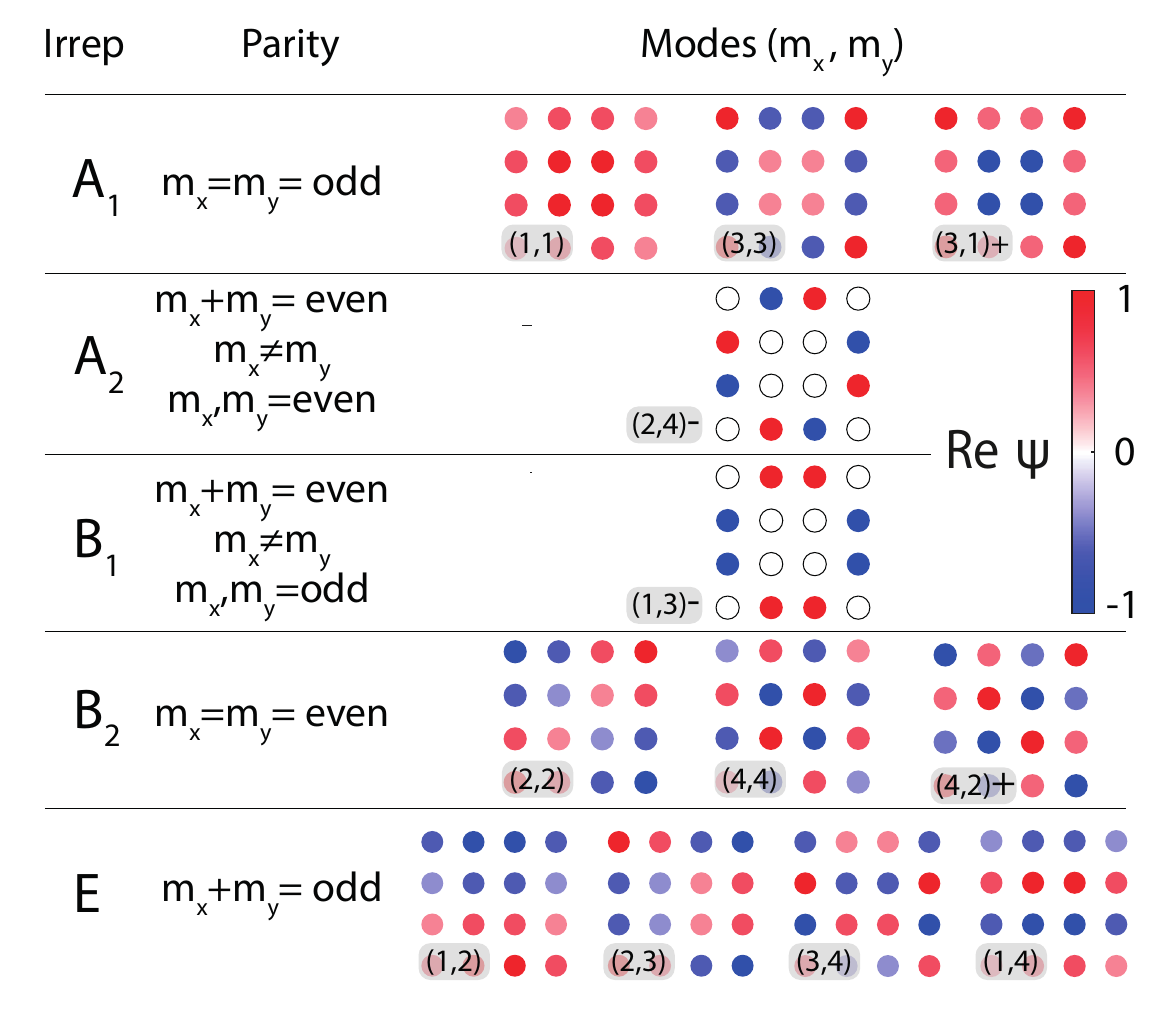}
    \caption{The symmetry classification of the basis functions $\psi^{(m_x,m_y)}$ with respect to the irreducible representations  and parity numbers $m_x$ and $m_y$ [see Eq.~\eqref{SW_expansion}]. Color shows the real part of the wavefunction amplitude at each lattice site.}
    \label{fig:2_table}
\end{figure}

The eigenstates of the finite lattices can be classified and characterized based on their distribution in the quasi-reciprocal space~\cite{Kornovan2016,Asenjo-Garcia2017Aug,Sheremet2023Mar} which can be done by expanding the dipole moments distribution of the $j$-th eigenstate $\Psi^{(j)}$ in the standing (Bloch) waves  basis:
\begin{multline} 
\Psi^{(j)}(n_x, n_y) = \sum_{m_x,m_y} c^{(j)}_{m_x,m_y} \psi^{(m_x,m_y)}  =  \\ = \dfrac{2}{N+1} \sum_{m_x,m_y} c^{(j)}_{m_x,m_y} \sin(q_0 m_x n_x) \sin(q_0 m_y n_y),
\label{SW_expansion}
\end{multline}
where $c^{(j)}_{m_x,m_y}$ are the expansion coefficients, $q_0 = \pi/(N+1)$ is the discretization quasi-wavevector, $m_{x,y} = 1..N$   are the numbers of the basis state in the quasi-reciprocal space, and $n_{x,y} = 1..N$ are the atoms positions in the array in the real space. {Such basis is straight-forwardly constructed as a tensor product of the two sets of standing waves in $x$ and $y$ directions.} Two indices, $m_x$ and $m_y$, that specify the basis state $\psi^{(m_x,m_y)}$, show that the standing wave distribution of the dipole moments changes the sign $m_x-1$ and $m_y-1$ times in $x$ and $y$ directions, respectively.  For example, the basis state that is closest to the corner of the BZ [$M$-point in Fig.~\ref{fig:fig1_scheme}(b)] in the reciprocal space is characterized by the indices $m_x=m_y=N$.

Alternative natural classification of the eigenstates is based on their symmetry properties. The square array belongs to the $C_{4v}$ point symmetry group and thus the eigenstates $\Psi^{(j)}$ should transform according to one of the irreducible representations (irreps) $A_1$, $A_2$, $B_1$, $B_2$, $E$. However, the basis given by Eq.~\eqref{SW_expansion} does not account for the specific symmetry of the square array. Therefore, not all of the basis states $\psi^{(m_x,m_y)}$ fall into the irreps of the $C_{4v}$ point symmetry group and, thus, should be symmetrised. Then, the following  correspondence between the basis states and irreps, summarized in Fig.~\ref{fig:2_table}, can be established. The basis states with odd value of $(m_x+m_y)$ transform according to $E$ irreducible representation and have ``azimuthal numbers'' $\pm1$ (or their linear combination), i.e. phases of the neighbor corner dipoles differ by $\pi/2$. The basis states with $m_x=m_y$ transform according to $A_1$ ($B_2$) irreducible representation if $m_{x,y}$ is odd (even), and have ``azimuthal numbers'' $0$ ($2$), i.e. phases of the neighbor corner dipoles are the same (differ by $\pi$). If $m_x \ne m_y$ and $(m_x+m_y)$ is even the mode $\psi^{(m_1,m_2)}$ hybridises with the mode $\psi^{(m_2,m_1)}$ and they form two linear combinations:
\begin{equation}
\psi^{(m_1,m_2)^{\pm}} = \dfrac{1}{\sqrt{2}} \left( \psi^{(m_1,m_2)} \pm \psi^{(m_2,m_1)} \right).
\label{eq:symm_basis}
\end{equation}
Antisymmetric modes transform according to $A_2$ or $B_1$, while symmetric modes transform according to $A_1$ or $B_2$ irrep. 
\subsection{Mechanism of the radiation suppression}

If the eigenstate $\Psi^{(j)}$ of the atomic array is mainly characterized by a single dominant contribution of a basis state $\psi^{(m_x,m_y)}$, its radiative losses $\Gamma_j$ are qualitatively determined by its {\it location in the BZ}, which defines the efficiency of the destructive interference between the neighbouring dipoles in the array. The losses are typically weaker for the states that are closer to the $M$-point in the reciprocal space, i.e. when $m_x,m_y \rightarrow N$, due to the larger mismatch between the wavevector of such state and the free-space wavevector.

The second factor affecting the radiative losses is associated with the {\it symmetry of the structure} and the eigenstates. For instance, Bloch waves propagating in in an infinite square lattice in directions near the $M$-point and symmetric with respect to the $\Gamma M$-direction are degenerate owing to the symmetry of the lattice, see Fig.~\ref{fig:fig1_scheme}. However, in a finite array the degeneracy is lifted due to the {\it external coupling} of these waves according to Friedrich-Wintgen mechanism~\cite{Friedrich1985a} leading to the formation of symmetric and antisymmetric states. The latter ones transform according to $A_2$ or $B_1$ irreps and are antisymmetric with respect to the diagonal vertical planes of symmetry $n_x = \pm n_y$, i.e. the excitation of the corner dipole moments of such states is necessarily zero. Due the suppressed scattering from the sharp edges of the structure the radiative losses of these states can substantially decrease~\cite{Asenjo-Garcia2017Aug,Poddubny2020} \cite{WiersigPRL2006}. On the opposite, symmetric modes that transform according to $A_1$ or $B_2$ with non-zero corner dipole moments are characterized with the increased losses. \\


\begin{figure*}[t!]
    \centering
    \includegraphics[width=1\textwidth]{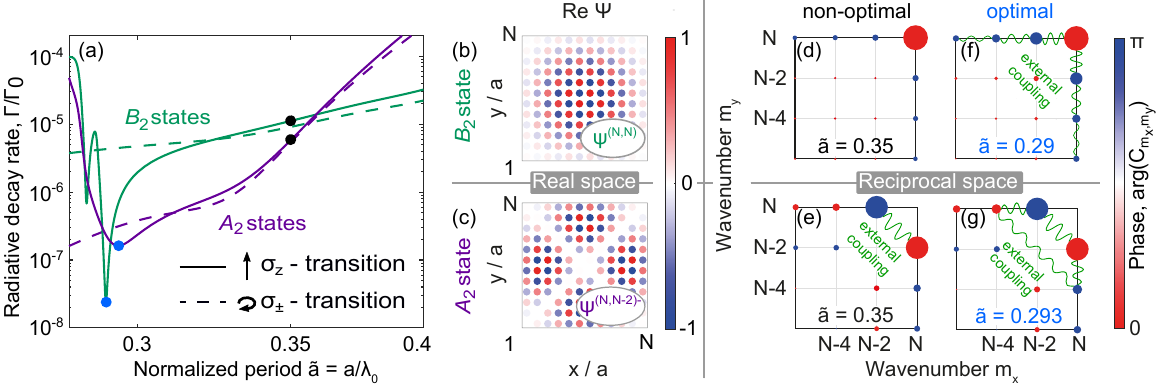}
    \caption{(a) Normalized radiative decay rate of two most subradiant eigenstates corresponding to $B_2$ (green curves) and $A_2$ (violet curves) symmetry in $12 \times 12$ square array as a function of the normalized period $\tilde{a}$. (b,c) Real part of the wavefunction plotted in real space for the states shown in (a) with green and violet solid curves, respectively. (d-g) Expansion of the eigenstates, Eq.~\eqref{SW_expansion}, shown in (a) with (d,e) black and (f,g) blue points; size of the circles indicate the amplitude of the corresponding harmonics; wavy lines illustrate the coupling between different harmonics.}
    \label{fig:Finite}
\end{figure*}
Additional mechanism of the losses suppression may appear due to the external coupling of the eigenstates unrelated to the symmetry of the structure, but rather to the accidental degeneracy of the Bloch waves in the corresponding infinite lattice. As we show further, such regime can be realised for small enough periods, when the lattice dispersion becomes {\it quasi-flat}. In this case two (or typically several) basis states interact via radiation continuum resulting in additional suppression or increase of their radiative losses. The overall radiative losses of an eigenstate are then determined as a result of the competition between these three mechanisms of loss suppression. In order to determine which of the eigenstates exhibits minimal losses for a given size and period of the structure, we perform straight-forward numerical calculations and the results are  presented in the next subsection.

\subsection{Numerical calculations of the finite array eigenstates}

In Fig.~\ref{fig:Finite}(a) we show numerically calculated emission rates $\Gamma/\Gamma_0$, normalized by the emission rate of a single atom, for two collective states of the $12 \times 12$ square array as a function of the normalized period $\tilde a=a/\lambda_0$. These states are characterized by the minimal emission rates among the others. Solid lines correspond to the  $\sigma_z$ dipoles, while the dashed lines to the $\sigma_{+}$ or $\sigma_{-}$ ones. One can observe that for periods $\tilde{a} \gtrsim 0.32$ the radiative losses monotonously increase with the decrease of the period. Standing wave decomposition Eq.~\eqref{SW_expansion} performed in this range of periods reveals that all states are characterized by the dominant contribution either of a single basis state or two basis states in the symmetrized form, Eq.~\eqref{eq:symm_basis}. The numerical analysis revealed that the most subradiant states (in the array with even $N$) transform according to either $B_2$ or to $A_2$ irrep. The distribution of the  wavefunction of the $\sigma_z$ polarized states in real space is shown in Figs.~\ref{fig:Finite}(b,c), while its representation in the reciprocal space, i.e. the decomposition Eq.~\eqref{SW_expansion}, for the period $\tilde{a}=0.4$ is shown in Figs.~\ref{fig:Finite}(d,e). The  distribution for the $\sigma_{\pm}$ look similar. As expected, the $B_2$ state is the closest one to the $M$-point and it is well approximated by the $\psi^{(N,N)}$ basis function. The dipole moments of this state in the neighbour atoms are almost out-of-phase, which leads to  their efficient destructive interference.

The radiative losses of such state, however, are minimal only for the periods $\tilde{a} \gtrsim 0.36$. For smaller periods, the $A_2$ state, which is the second closest to the $M$-point, becomes more subradiant. This state has the dominant components $m_x=N, m_y=N-2$ and $m_x=N-2, m_y=N$ in the antisymmetric combination $\psi^{(N,N-2)^-}$, Eq.~\eqref{eq:symm_basis} (see also Fig.~\ref{fig:fig1_scheme}). In accordance to the symmetry properties of the irrep $A_2$, the dipole moments of such state are equal to zero on diagonals. This leads to small probability that the excitation resides on the corners of the square, see Fig.~\ref{fig:Finite}(c), leading to suppressed scattering from sharp edges and consequently weak radiative losses of such state in the most range of the periods.

Such antisymmetric subradiant states can also be viewed as the result of the external coupling between two eigenstates of the rectangular array that appears under stretching/squeezing the lattice to the square one~\cite{YangIOP2016}, see also Supplementary Materials, Sec. A for the details. Consequently, the mechanism of the symmetry-induced external coupling is very sensitive to the deformation of the lattice: in a rectangular lattice the efficiency of the coupling drops very fast, resulting in the substantially increased losses of the $\psi^{(N,N-2)^{-}}$ state even for very small detunings from the square geometry of the order of 1\%. On the other hand, the $\psi^{(N,N)}$ state, not being associated with the specific symmetry of the structure, survives under strong deformation of the lattice up to tens of percents.

For even smaller periods $\tilde{a} \approx 0.3$ the decay rates of both $A_2$ and $B_2$ states for $\sigma_z$ polarization exhibit a minimum for a given number of atoms, see Fig.~\ref{fig:Finite}(a). At these minima the wavefunctions of the states in the real space remain similar to those in Figs.~\ref{fig:Finite}(b,c). However, the expansion in the reciprocal space, Eq.~\ref{SW_expansion}, shown in Figs.~\ref{fig:Finite}(f,g), reveals that the amplitudes of other harmonics are substantially increased. The interaction between these harmonics leads to additional suppression of radiative losses near the period $\tilde{a} \approx 0.3$. At the same time, the calculations for the $\sigma_{\pm}$ dipole transitions, dashed curves in Fig.~\ref{fig:Finite}(a), show that radiative losses are monotonous function of the period in this case and no resonant suppression of the losses via external coupling mechanism is observed.

\subsection{Infinite atomic lattices}
\begin{figure}[t]
    \centering
    \includegraphics[width=1\columnwidth]{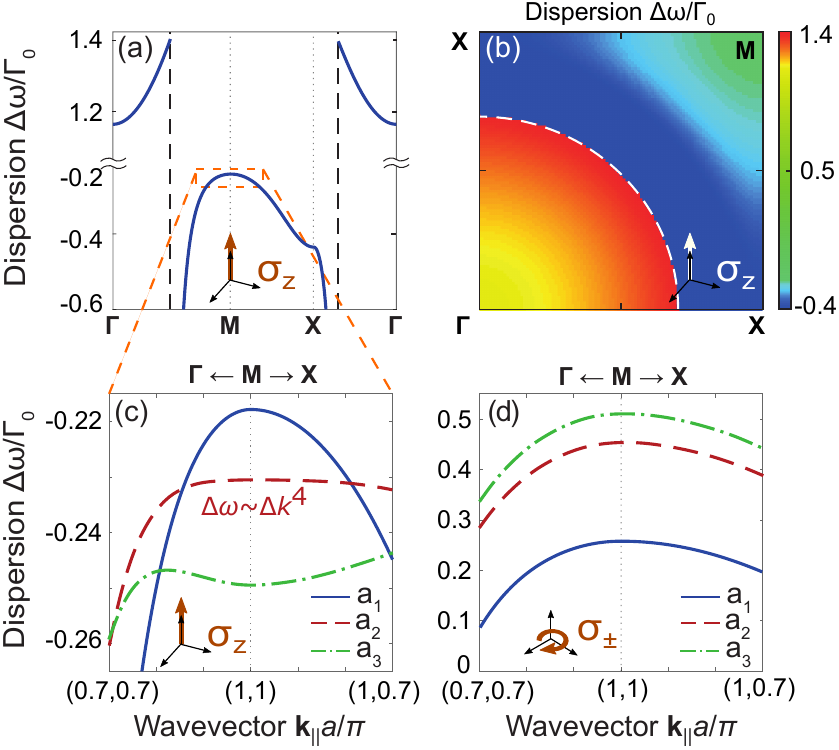}
    \caption{(a,b) Dispersion of an infinite square lattice of $\sigma_z$ polarized atoms for the period $\tilde{a} = 0.35$ (a) along high symmetry path and (b) in the whole BZ. (c,d) Dispersion near the $\text{M}$ point for periods $\tilde{a}_1 = 0.35$ (blue curves), $\tilde{a}_2 = 0.294$ (red curves) and $\tilde{a}_3 = 0.28$ (green curves) for (c) $\sigma_z$ polarized atoms and (d) $\sigma_{\pm}$ polarized atoms.}
    \label{fig:Dispersion}
\end{figure}

Appearance of the external coupling near $\tilde{a} \sim 0.3$ for $\sigma_z$ polarized dipoles and its absence for the $\sigma_{\pm}$ ones can be linked to the behaviour of the infinite lattice dispersion of the polaritonic states near the $M$-point. To illustrate this, we performed calculations of the band diagrams of the square lattice for both polarizations of the dipole transitions and for different periods, see details of the calculations in the Supplementary Materials, Sec. B. First, we consider the most interesting case of the $z$-oriented dipoles. In Fig.~\ref{fig:Dispersion}(a), we plot the dispersion along the main direction in the first BZ. Fig.~\ref{fig:Dispersion}(b) complements the panel Fig.~\ref{fig:Dispersion}(a) with the 2D colorplot in the whole BZ. Above the light cone, shown with the black dashed lines in Fig.~\ref{fig:Dispersion}(a) and with white dashed curve in Fig.~\ref{fig:Dispersion}(b), the Bloch states are leaky and the eigenstates of the corresponding finite arrays are characterized with large radiative losses. On the other hand, the modes below the light line are guided (the eigenfrequency is purely real) and the eigenstates of the corresponding finite arrays are subradiant.

As it was shown in the previous subsection, the eigenstates of the finite arrays that are closest to the $M$-point, are the most subradiant due to the largest mismatch with the wavevectors of the free space. One can observe in Fig.~\ref{fig:Dispersion}(c) that the dispersion near the $M$-point turns out to be qualitatively different for different periods. For the large periods $\tilde{a} \gtrsim 0.3$ the interaction between the dipoles is rather weak, so that the dispersion is parabolic. For the lattice period near $\tilde{a}=0.294$ the dispersion exhibits an inflection point and becomes quartic, as shown with the red curve in Fig.~\ref{fig:Dispersion}(c). For even smaller periods $\tilde{a} \lesssim 0.294$, the dispersion becomes nonmonotonous near the $M$-point along the $\Gamma M$ direction [green curve in Fig.~\ref{fig:Dispersion}(c)], which explains the external coupling of the several states as now their energies match with each other, analogously to the one-dimensional case~\cite{Kornovan2021Dec}. In contrast, dispersion in the case of $\sigma_{+,-}$ polarized dipole transition, shown in Fig.~\ref{fig:Dispersion}(d) with dashed curves, remains monotonic near the  $M$-point for any subwavelength period. This explains the absence of the external coupling in finite arrays for this polarization.

\section{Discussion}


\subsection{Variation of the radiative losses with the size of arrays with different geometries}

\begin{figure}[t!]
    \centering
    \includegraphics[width=1\columnwidth]{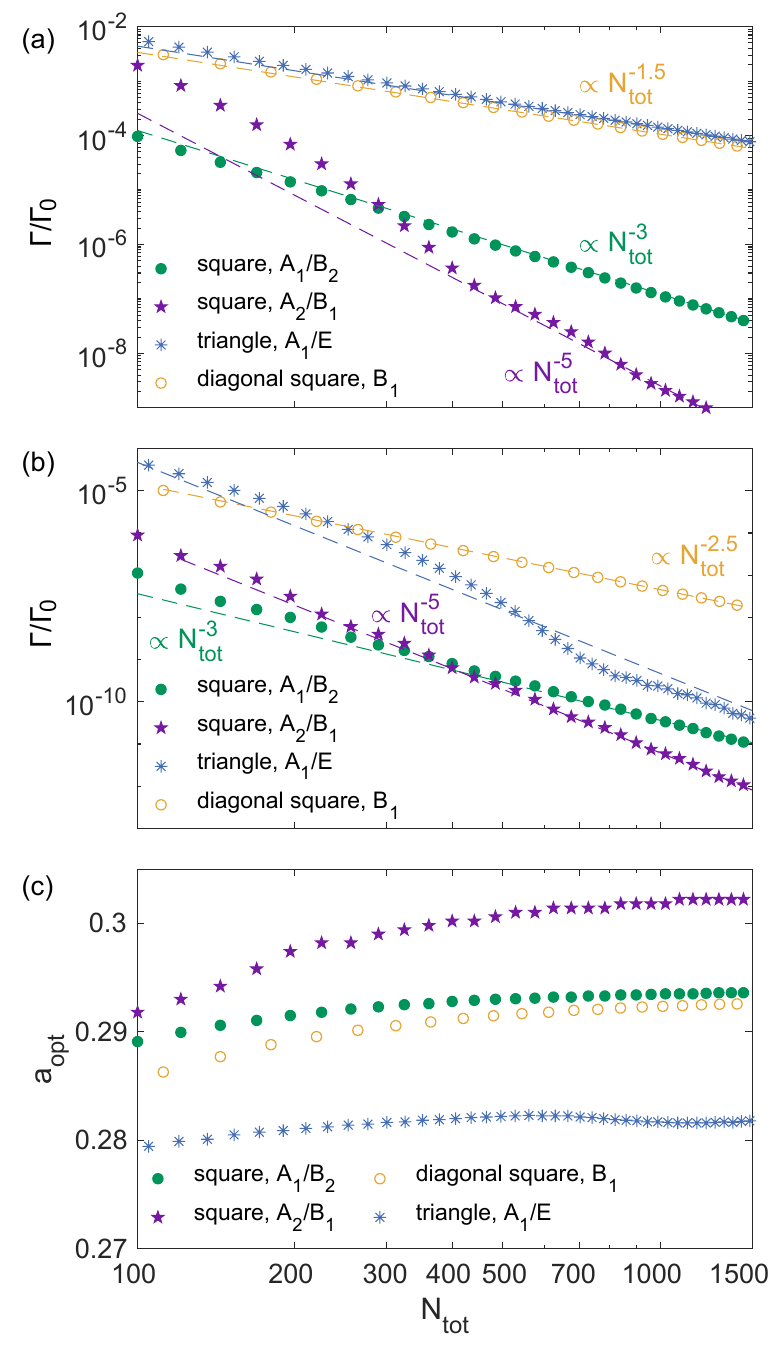}
    \caption{(a,b) Radiative losses of subradiant states as a function of the total number of atoms in the array $N_{tot}$ for different geometries for (a) fixed distance between the neighbor atoms $\tilde{a} = 0.4$, (b) optimal period for each $N_{tot}$, shown in panel (c). Dashed lines indicate corresponding polynomial functions.}
    \label{fig:Scaling}
\end{figure}

The mechanisms that determine the formation of the subradiant states in atomic arrays shown on the example of the square geometry have rather general nature and are present in various configuration of atomic ensembles. Calculations performed for a few other main geometries of the regular atomic arrays such as rotated square, triangle, hexagon (see Appendix, Sec C. for the details) have revealed that the square geometry is preferable from the point of view of the state lifetime. To illustrate this, in Fig.~\ref{fig:Scaling}(a) we show decay rate of the four characteristic eigenstates (corresponding to different geometries of the array) as a function of the total number of the atoms in the array $N_{tot}$, calculated for the non-optimal period $\tilde{a} = 0.4$ when there is no accidental external coupling between the eigenstates caused by the quasi-flat dispersion. The polarization of the dipoles was chosen to be along the $z$ direction. The $N_{tot}^{-5} \equiv N^{-10}$ scaling law of the $A_2/B_1$ state in a square array, shown with purple pentagrams in Fig.~\ref{fig:Scaling}(a), turns out to be unique. For instance, diagonal square or triangle arrays exhibit only $\propto N_{tot}^{-1.5}$ scaling laws providing much shorter lifetimes than the square geometry even for rather small $N_{tot}$.

Another type of the eigenstate of the square array, $A_1/B_2$ states with the dominant $\psi^{(N,N)}$ contribution, exhibit only $\propto N_{tot}^{-3}\equiv N^{-6}$ dependence of the decay rate. Such difference in the scaling laws for this particular geometry can be qualitatively explained in a simple manner, as elaborated in the Supplementary Materials, Sec. D. Despite the slower decrease of the decay rate for $A_1/B_2$ states, such asymptotic dependence is valid only for large number of atoms, while it does not necessarily reflect the difference between the radiative losses for a particular $N_{tot}$. One can observe in Fig.~\ref{fig:Scaling}(a), that the antisymmetric state becomes the most subradiant only for $N_{tot}\gtrsim300$ for a given period $\tilde{a}=0.4$.

In Fig.~\ref{fig:Scaling}(b) we plot the calculated decay rates for the optimal periods $\tilde{a}_{opt}$, for which the losses are minimal (i.e. period is optimized for each value of $N_{tot}$), as indicated in Fig.~\ref{fig:Scaling}(c). Here, the presence of external coupling leads to a change of the scaling laws in some geometries while remaining the same $\propto N_{tot}^{-3}$ and $\propto N_{tot}^{-5}$ in the square geometry. This contrasts with the one-dimensional case, where interaction between two standing waves leads to the decrease of the asymptotic behaviour of the subradiant state decay rate from $\propto N^{-3}$ to $\propto N^{-7}$~\cite{Kornovan2019Dec,Kornovan2021Dec}. Although, e.g. in a triangle geometry decrease of the decay rate also becomes close to $\propto N_{tot}^{-5}$, the absolute values of the decay rate still remain minimal for the $A_2/B_1$ states of the square geometry.

\subsection{Excitation of the subradiant states}

In order to demonstrate the possibility of excitation of the strongly subradiant states appearing for the $\sigma_z$ polarization, we consider the scattering of vector Bessel beams~\cite{Richards1959Dec, Kotlyar2019Oct} by a square atomic array. The use of such specific excitation field has two underlying reasons. First, we are considering the states characterized by large wavevectors. The use of Bessel beams, which can have arbitrary angular momentum, allows to partially match the azimuthal component of the wavevector of the incident field to that of the eigenstate. Second, Bessel beams have non-zero longitudinal electric field, which allows for efficient excitation of the $\sigma_z$ polarized states. The considered highly inhomogeneous incident field may excite various modes of the lattice with different efficiency, therefore we have optimized orbital number of the beam $m$ to maximize the overlap between external electric field and the desired collective state, see the Supplementary Materials, Sec. E for details of the scattering calculations).

\begin{figure}[t!]
    \centering
    \includegraphics[width=1\columnwidth]{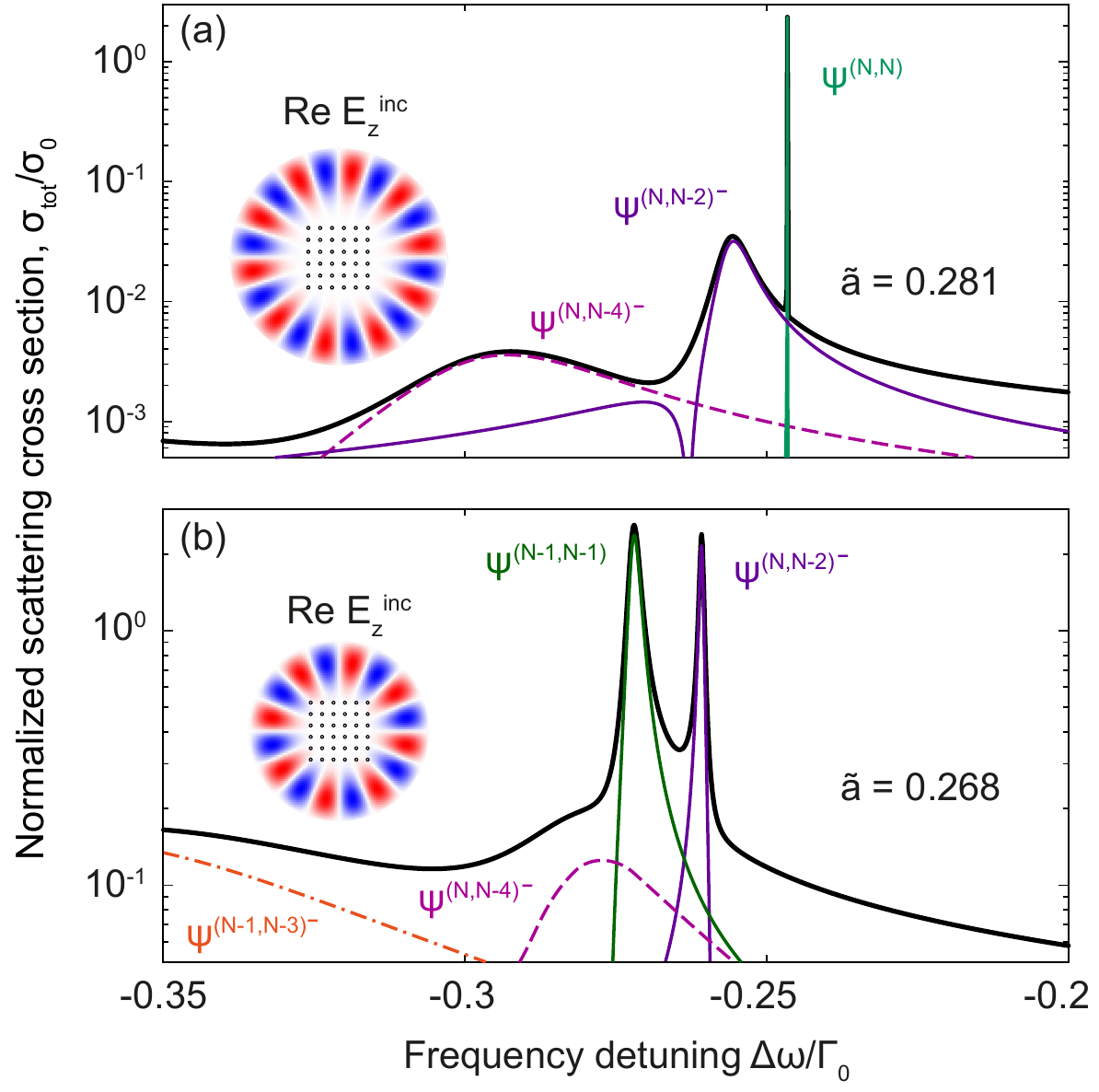}
    \caption{Normalized scattering cross section spectra of 6$\times$6 square atomic lattice irradiated by Bessel beam with (a) azimuthal number $m=9$ and period $\tilde{a}=0.281$ and (b) $m=7$ and $\tilde{a}=0.268$. Solid black curves correspond to the total scattering cross section spectra, while colored curves correspond to  different eigenstates with dominant contributions of basis states indicated in the plots. Insets show distributions of $z$ component of the incident electric field. }
    \label{fig:Scattering}
\end{figure}

In Fig.~\ref{fig:Scattering}, we show numerically calculated scattering cross-section spectra of $6\times 6$ square atomic array irradiated by normally incident vector Bessel beam with a spin $s=1$ due to circular polarization, and orbital numbers $l=7$ or $l=9$ \cite{Kotlyar2019Oct}. Total cross-section consists of various peaks matching eigen modes of the lattice characterized with different dominant basis states. The parameters of the beams were chosen to achieve the maximum coupling with the eigen states with dominant contribution of basis states $\psi^{(N,N-2)^-}$ and $\psi^{(N,N)}$ respectively. Lattice periods correspond to optimal periods in terms of radiative losses of these states,  $\tilde{a} = 0.268$ and $\tilde{a} = 0.281$ in  the case of $6\times 6$ array. Cross-section spectra are normalized with a maximum value of a single atom cross-section $\sigma_0$.

As one can see in Fig.~\ref{fig:Scattering} insets, the optimal incident field profile barely overlaps the atomic array from the outside, setting $\pi$ phase shift between neighboring atoms at the edges with almost no influence at internal atoms. Therefore, a total angular momentum of the beam $l+s$ matches to either a half of total number of edge atoms $2(N-1)$, see Fig.~\ref{fig:Scattering}(a), or the same number but excluding corner atoms $2(N-2)$, see Fig.~\ref{fig:Scattering}(b). In the first case, the beam is capable to excite the $B_2$ state with a dominant $\psi^{(N,N)}$ contribution. In the second case, where the field at the corners is absent, it is optimal to excite $A_2$ state with a dominant $\psi^{(N,N-2)^-}$ contribution. Thereby a fine tuning of the incident beam frequency and angular momentum allows to selectively excite the desired long-living eigen mode of the array.

\subsection{Relation to high quality modes in nanophotonic cavities}

Importantly, the generality of the reported results is also underlined by the tight connection to the  research area of engineering of planar dielectric nanophotonic cavities, such as homogeneous dielectric cavities~\cite{GuoIEEE2003,LiIEEE2006,YangIEEE2007,YangPRA2007,BittnerPRE2013,CaoRevModPhys2015,YangIOP2016} and periodic or quasi-periodic nanostructured cavities~\cite{TsiaOE2004,NohPRA2010}.

The resonators made of homogeneous dielectric are expected to have the highest $Q$-factors for  spherical or cylindrical geometries, which exhibit the highest symmetry. However, the utilized fabrication methods or proposed technological applications sometimes demand the use of the cavities of the other shapes, e.g. cubes, squares, hexagons etc., which were also shown to support rather high $Q$-factors and have potential in development of micro and nanolasers~\cite{YangIOP2016,ZhengZhengACSNano2018,TiguntsevaACSNano2020}. The dispersion of homogeneous dielectric is, however, quasi-linear and the external coupling due to quasi-flat band in this case is not possible. Unlike the homogeneous cavitites, the dispersion in the nanostructured ones, i.e. photonic-crystal or nanoparticle cavities, is more tunable. Consequently, quasi-flat and nonmonotonous bands in such structures result in the quality factor boost in the finite arrays~\cite{NohPRA2010}. By proper translating all mechanisms of the subradiant states formation reported in this work for the dipolar arrays onto the nanophotonic platform could lead to the development of the novel designs of the compact optical micro and nanocavities.

\section{Summary}
To summarize, we have studied main factors that define the radiative loss suppression in finite-size planar atomic arrays. We have shown that the square arrays support eigenstates with minimal losses among different regular arrangements of the atoms, that scale with the number of atoms as $\propto N_{tot}^{-5}$. The two distinct types of external coupling associated with the symmetry of the array and with the quasi-flat dispersion of the corresponding infinite array play the dominant role in the strong boost of the radiative lifetime of such states. We believe, our findings might provide useful insight in designing of the cold atoms based qubit arrays for storing and coherent manipulation of quantum information.

\section{Acknowledgement}
The work was supported by the Federal Academic Leadership Program Priority 2030. The numerical computations were supported by the Russian Science Foundations grant No. 21-72-10107. We thank Yuri Kivshar and Ivan Iorsh for significant discussions.

\bibliographystyle{ieeetr}
\bibliography{ref}
\end{document}


\title{Strongly subradiant states in planar atomic arrays:\\Supplementary Material}




\maketitle
\tableofcontents

\section*{A. Effect of deformation of the square array on the eigenstates decay rates}

\begin{figure}[t]
    \centering
    \includegraphics[width=0.7\textwidth]{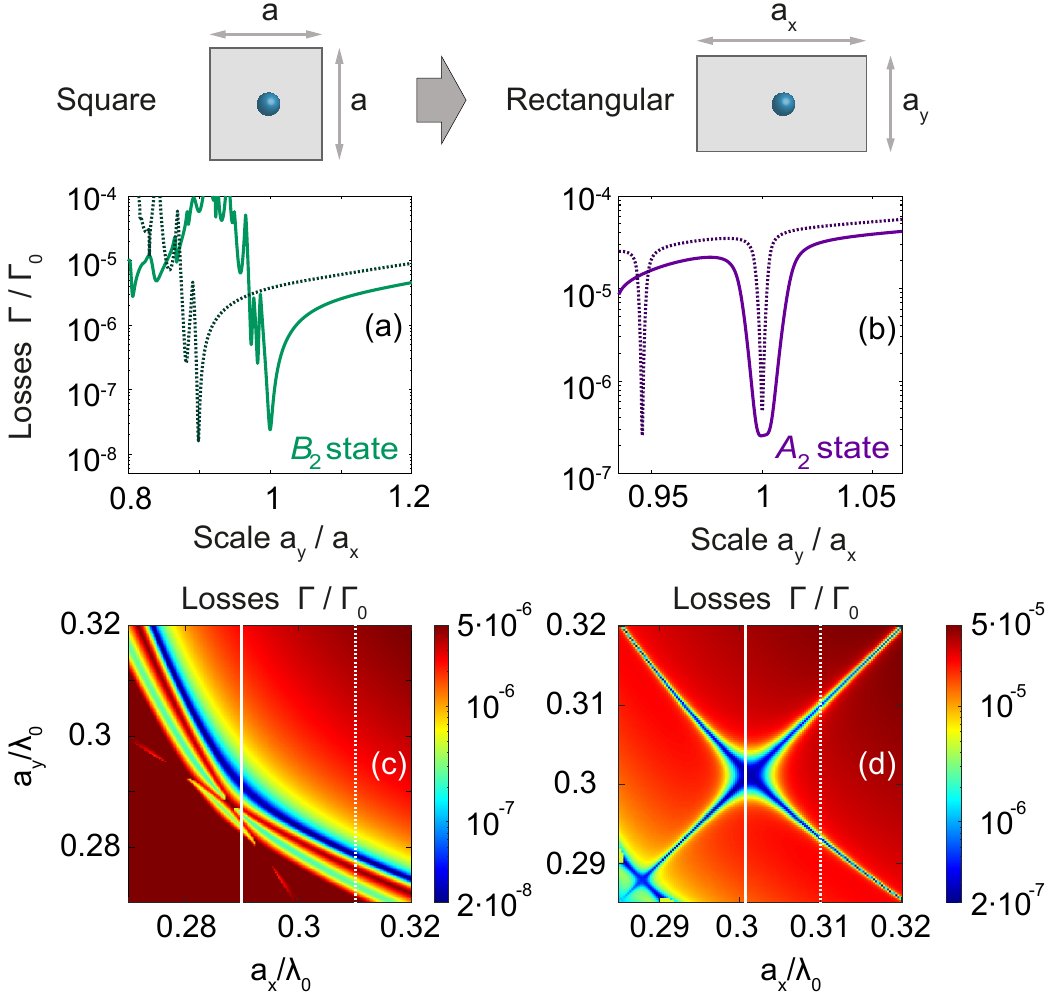}
    \caption{Normalized radiative decay rate of most subradiant $B_2$ (a,c) and $A_2$ (b,d) states in rectangular $12 \times 12$ array.  In (a) and (b) period $\tilde{a}_{x}$ along $x$-axes is fixed whereas period $\tilde{a}_{y}$ along $y$-axis is changing. In colormaps (c) and (d) both periods $\tilde{a}_{x}$ and $\tilde{a}_{y}$ are varied.  In (a) and (c) solid lines correspond to $\tilde{a}_{x} = 0.29$, dotted lines --- to $\tilde{a}_{x} = 0.31$, whereas in (b) and (d) solid lines correspond to $\tilde{a}_{x} = 0.301$, dotted lines --- to $\tilde{a}_{x} = 0.31$.}
    \label{fig:rectangular}
\end{figure}

In this section, we analyze sensitivity of the radiative decay rate of the two most subradiant states in the square atomic array (see Fig.~3 in the main text) to the deformation of the lattice. To illustrate this, we have calculated decay rates of the two most subradiant states in the square $12\times 12$ array arranged in a rectangular lattice with periods $\tilde{a}_x, \tilde{a}_y$ as a function of the ratio of the periods $a_y/a_x$. The results of the calculations for the period $\tilde{a}_x=0.31$ are shown in Figs.~\ref{fig:rectangular}(a,b) with dotted curves. We should note that in the case of rectangular lattice, which belongs to the point symmetry group $C_{2v}$, all the eigenstates are divided into four symmetry types depending on their transformation with respect to the two vertical symmetry planes. On the other hand, in the case of square lattice (with equal number of atoms along $x$ and $y$ directions), which belongs to the $C_{4v}$ point symmetry group, there are five type of states. Unlike the square lattice, in the rectangular lattice all of the states with the dominant contribution of the $\psi^{(N,N)}$, $\psi^{(N,N-2)}$ or $\psi^{(N-2,N)}$ harmonic (see Eq.~(2) in the main text) fall into the same irreducible representation. In Figs.~\ref{fig:rectangular}(a,b) eigenstates with the dominant contribution of the $\psi^{(N,N)}$ and $\psi^{(N-2,N)}$ harmonics, respectively, are shown.


For large enough periods and not very large anisotropy of the lattice, the state with dominant contribution of $\psi^{(N,N)}$ harmonic does not interact with any other state upon the change of the $a_y/a_x$ ratio around $a_y/a_x=1$, since this state is not associated with the specific symmetry of the square lattice. Consequently, the radiative decay rate in Fig.~\ref{fig:rectangular}(a) shown with the point curve changes only slightly even for relatively large detunings of the $a_y/a_x$ from one up to several percents. On the other hand, the state with dominant contribution of $\psi^{(N-2,N)}$ harmonic hybridizes with the $\psi^{(N,N-2)}$ harmonic when the periods are close to be equal, i.e. the lattice is almost square. Such symmetry-induced external coupling leads to the strong suppression of losses of such state in the square lattice. Reciprocally, one can say, that the antisymmetric states in the square lattice are rather sensitive to the deformation, resulting in the substantially increased losses even for a very small detunings $\Delta (a_y/a_x) \lesssim 1\%$ from the square geometry, see Fig.~\ref{fig:rectangular}(b), point curve.

For even larger detuning ($a_y/a_x \approx 0.9$ ($\approx 0.94$) for $B_2$ ($A_2$) state for the given period $\tilde{a}_x=0.31$) there appears additional external coupling between with the $\psi^{(N,N-2)}$ harmonic caused by the quasi-flat band dispersion of the lattice. This also leads to substantial suppression of losses with the same level as in the square lattice with optimal period, shown with solid curves in Figs.~\ref{fig:rectangular}(a,b). Fig.~\ref{fig:rectangular}(c,d) summarizes the effects of stretching/squeezing of the square lattice on the decay rates of the subradiant states.


\section*{B. Calculation of the infinite lattice dispersion}

Here, we provide the details of the calculations of the dispersion of the various two-dimensional dipole lattices in a free space. According to the Bloch theorem, it is possible to present a eigenstate of an arbitrary Bravais lattice as $|\psi_\mathbf{k} \rangle = \sum_j e^{i\mathbf{k\cdot R}_j} \sigma_j |0\rangle$, where $\mathbf{R}_j$ is a position vector of atom $j=1\ldots \infty$. Such Bloch state is fully defined by the quasi-momentum $\mathbf{k}$, which can always be chosen to be within the first Brillouin zone. Plugging a Bloch state into the Hamiltonian, Eq.~(1) from the main text, for $N \rightarrow \infty$ and multiplying both sides of it by $\langle \psi_\mathbf{k}^\dag |$, we obtain the following general relations for the eigenfrequency and the decay rate~\cite{Asenjo-Garcia2017Aug}:


\begin{equation}
  \label{General dispersion}
  \begin{aligned}
    &\frac{\Delta \omega (\mathbf{k})}{\Gamma_0} = - \frac{3\pi}{k_0} \text{Re} [C(\mathbf{k})] ,\\       
    &\frac{\Gamma (\mathbf{k})}{\Gamma_0} = 1 - \frac{6\pi}{k_0} \text{Im} [C(\mathbf{k})],
  \end{aligned}
\end{equation}
where $\mathbf{R}_{ij} = \mathbf{R}_{j} - \mathbf{R}_{i}$, $C(\mathbf{k}) = \sum_{i\neq j} \mathbf{e_d^*} \cdot\mathbf{G}(\mathbf{R}_{ij})\cdot \mathbf{e_d} \ e^{i\mathbf{k\cdot} \mathbf{R}_{ij}}$, $\mathbf{e_d}$ is the unit vector parallel to the dipole moment transition, and $\mathbf{G}(\mathbf{R}_{ij})$ is the electromagnetic Green's tensor of a free space~\cite{Novotny2012Sep}: 
\begin{multline}
    \mathbf{G}(\mathbf{R},\omega_0) = \frac{ik_0}{6\pi} \delta^{(3)}(\mathbf{R})\hat{\mathbf{I}}+ \\ + \frac{e^{i k_0 R}}{4\pi R} \left[ \left(1+\frac{i}{k_0R}-\frac{1}{k_0^2R^2})\hat{\mathbf{I}} + (1-\frac{3i}{k_0 R}-\frac{3}{k_0^2R^2} \right) \frac{\mathbf{R}\otimes\mathbf{R}}{R^2}\right],
\label{Green}
\end{multline}
where $k_0 = \omega_0/c$ is a vacuum wavevector and $\hat{\mathbf{I}}$ is the unit dyadic. 

Note that infinite series in Eqs.~\eqref{General dispersion} have convergence issues due to the farfield term $e^{ik_0 R}/R$ in Eq.~\eqref{Green}. To overcome this problem, we exploit the Poisson summation approach, which was fully described for a rectangular lattice consisted of in-plane dipoles in Ref.~\cite{Belov2005Aug}. For a case of the $z$-polarised dipoles, one can step-by-step follow the derivation procedure in the Ref.~\cite{Belov2005Aug}, with a straight-forward substitution of the Green's tensor component $G_{zz}$ instead of $G_{xx}$ and obtain a dipole sum:
\begin{multline}
C(\mathbf{k}) = \frac{1}{4\pi a_x} \sum_{\varepsilon = \pm 1} \left[  \text{Li}_1 (e^{ia_x(k_0+\varepsilon k_x)}) +\frac{i}{k_0 a_x} \text{Li}_2 (e^{ia_x(k_0+\varepsilon k_x)}) - \frac{1}{k_0^2 a_x^2} \text{Li}_3 (e^{ia_x(k_0+\varepsilon k_x)}) \right] + \\ + \frac{1}{\pi a_x k_0^2} \sum_{Re [p_m] \neq 0} \sum_{n=1}^{+\infty} \left[k_0^2 K_0(p_m a_y n) - \frac{p_m}{a_y n} K_1(p_m a_y n) \right] \text{cos}(k_y n a_y) + \\ + \frac{1}{2\pi a_x k_0^2} \sum_{Re [p_m] =0} \biggl\{ (k_0^2+\frac{p_m^2}{2})\left[ \text{ln} (\frac{|p_m|a_y}{4\pi})+i\frac{\pi}{2}+\gamma \right] - \frac{p_m^2}{4}-\frac{k_y^2}{2} - \frac{\pi^2}{3a_y^2} +\\+ \frac{\zeta(3) a_y^2}{32\pi^2} \left[ 4k_0^2(2k_y^2-p_m^2) + p_m^2 (4k_y^2-p_m^2)\right] + \frac{\pi}{a_y}\left[ ik_z^{(m,0)}+\frac{k_0^2}{i k_z^{(m,0)}} \right] + \\ + \sum_{n=1}^{+\infty} \frac{\pi}{a_y} \left[ \frac{q^2}{ik_z^{(m,n)}} + \frac{k_0^2}{ik_z^{(m,-n)}} + ik_z^{(m,n)} + ik_z^{(m,-n)} \right] - \frac{4\pi^2 n}{a_y^2}-\frac{k_0^2}{n}-\frac{p_m^2}{2n}- \\ -\frac{a_y^2}{32\pi^2 n^3} \left[ 4k_0^2(2k_y^2-p_m^2) + p_m^2 (4k_y^2-p_m^2)\right] \biggr\},
\label{Dipole_sum_z}
\end{multline}
where:
\begin{equation*}
p_m = \sqrt{(k_x^{(m)})^2 - k_0 ^2}, \ \ k_x^{(m)} = k_x +\frac{2\pi m}{a_x},
\end{equation*}
\begin{equation*}
i k_z^{(m,n)} = \sqrt{(k_x^{(m)})^2 + (k_y^{(n)})^2 - k_0^2}, \ \ k_y^{(n)} = k_y +\frac{2\pi n}{a_y}.
\end{equation*}
In Eq.~\eqref{Dipole_sum_z},  $\text{Li}_n$ is a polylogarithm of order $n$, and $\text{K}_m$ is a Macdonald function of order $m$.
Also, here we use specific mathematical constants, such as Euler's constant $\gamma \approx 0.577$ and value of Riemann zeta function $\zeta(3) \approx 1.202$.


To obtain analogous dipole sum for the dipole with circular polarized transition moment in a rectangular lattice, one should substitute $\mathbf{e_d}$ with the $\mathbf{e_\pm} = \dfrac{1}{\sqrt{2}} (\mathbf{e_x} \pm i\mathbf{e_y})$, which results in the change of $G_{zz}$ to the $\dfrac{1}{2} (G_{xx}+G_{yy})$. Therefore, one can directly use find expressions (A19, A22) obtained in Ref.~\cite{Belov2005Aug} for $G_{xx}$ ($x$-oriented dipoles), then apply it for $G_{yy}$ ($y$-oriented dipoles) with a permutation of $x$ and $y$. Finally, taking a half sum for two dipole sums, one may get the final answer.

The obtained infinite series can be accurately truncated and calculated numerically due to fast convergence of order $1/n^4$~\cite{Belov2005Aug}. After plugging the obtained dipole sums into Eq.~\eqref{General dispersion}, taking certain parameters of a square lattice, one may get dispersion dependencies $\Delta \omega (\mathbf{k})$ in the main text, see Fig.~1(a) and Fig.~4.

\section*{C. Subradiant states in atomic arrays with different geometry}

\begin{figure}[t]
    \centering
    \includegraphics[width=1\columnwidth]{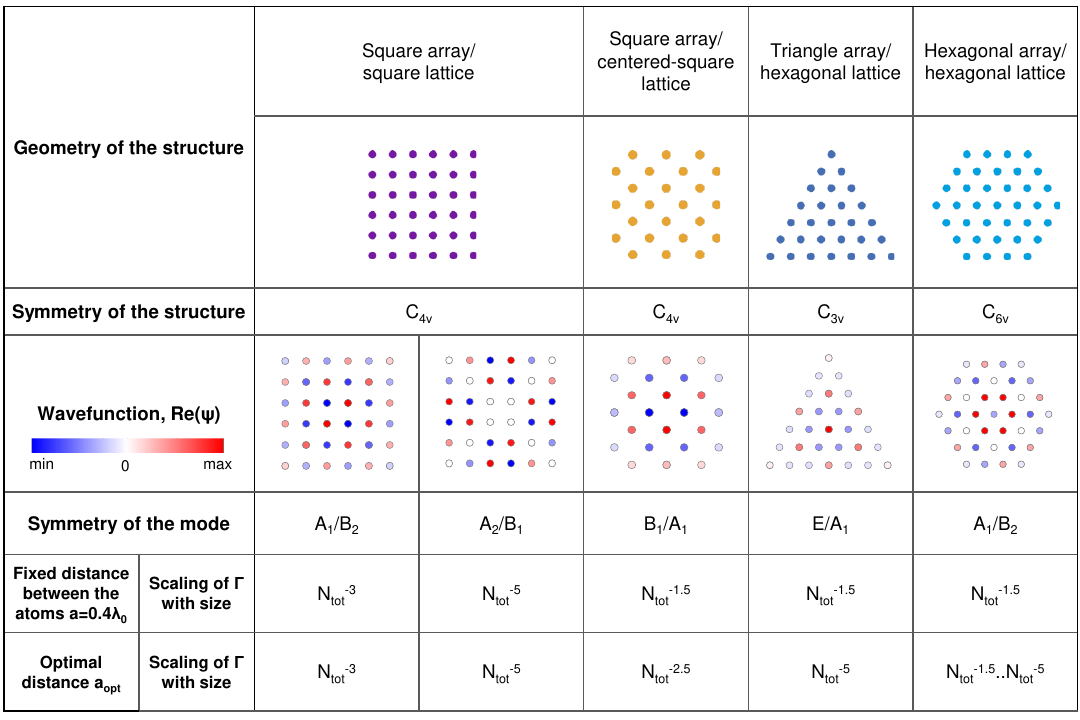}
    \caption{Table containing the schematics of the considered arrays and their main characteristics.}
    \label{fig:table}
\end{figure}

We have numericaly studied subradiant states in regular arrays of different geometry arranged in square or hexagonal lattices. Here, we present the results of numerical simulations for the considered cases: (i) square array arranged in square lattice cut out along the unit vectors, (ii) square array arranged in diagonal square lattice, (iii)/(iv) triangle/hexagonal array cut out of the hexagonal lattice. Such arrays belong to $C_{4v}$, $C_{4v}$, $C_{3v}$ and $C_{6v}$, respectively. The schematics of the structures are given in the Fig.~\ref{fig:table}. Note that the geometry (ii) can be viewed as the square cut out along the diagonals of the simple square lattice.

The considered square and hexagonal infinite lattices both exhibit quasiflat-band dispersion for the period $\tilde{a} \approx 0.3$ for the case of $z$ oriented dipole transition moments of the emitters (not shown here). Consequently, we distinguish two different regimes similarly to the main text. First one, when the normalized lattice period $\tilde{a}$ is larger than $\approx 0.3$; in further simulations we fix the period as $\tilde{a}=0.4$ in this regime. The second one, corresponds to $\tilde{a} \lesssim 0.3$, when additional external coupling due to quasiflat-band dispersion appears. In this regime, in the calculations we optimize the period for each value of $N_{tot}$ to achieve the minimal decay rate.

\begin{figure}[t]
    \centering
    \includegraphics[width=0.56\textwidth]{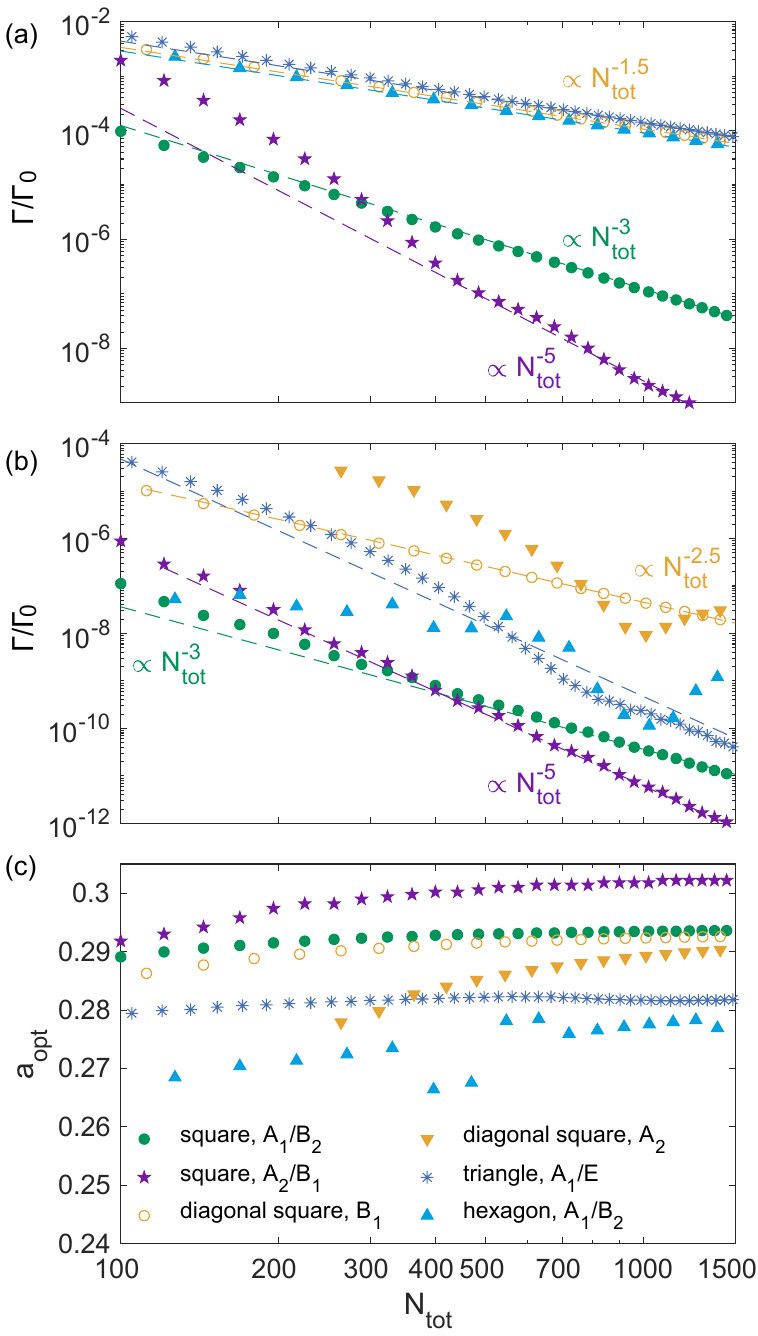}
    \caption{(a,b) Radiative losses of subradiant states as a function of the total number of atoms in the array $N_{tot}$ for different geometries illustrated in Fig.~\ref{fig:table} for (a) fixed distance between the neighbor atoms $\tilde{a} = 0.4$, (b) optimal period for each $N_{tot}$, shown in panel (c). Dashed lines indicate corresponding polynomial functions. The legend in (c) is valid for all panels.}
    \label{fig:scaling_supp}
\end{figure}

We numerically calculated the radiative decay rate as a function of $N_{tot}$ for all eigenstates of all considered arrays, and for two above-mentioned regimes. For each geometry we have identified the symmetry of the most subradiant eigenstates (in the limit of large number of atoms in array $N_{tot}$) and plotted the wavefunction of this state in an array of few atoms for the illustrative purposes, see Fig.~\ref{fig:table}, third row. The radiative decay rate of the most subradiant states as a function of $N_{tot}$ is shown in Fig.~\ref{fig:scaling_supp}(a). The calculations revealed that for the large $N_{tot}$ the states with the $A_2/B_1$ symmetry in the square array are the most subradiant ones among the considered geometries for any period. For large periods they exhibit unique $N_{tot}^{-5}$ decrease of the radiative losses with the size of the array, while other geometries show much slower $N_{tot}^{-1.5}$ dependence. 

For smaller periods the size dependencies remain the same in the square arrays, while they change in the other geometries, Fig.~\ref{fig:scaling_supp}(b). Moreover, some of the states start to exhibit slightly non-monotonic dependence, see e.g. $A_2$ mode of the diagonal square array in Fig.~\ref{fig:scaling_supp}(b). Surprisingly, the triangle array with the sharpest corners supports eigenstates with dependence close to $N_{tot}^{(-5)}$. For not so large number of atoms $N\lesssim 100$, however, the size scaling rate becomes even less specific and the decay rate of the hexagon eigenstates can slightly exceed those of the square array ones.

\section*{D. Qualitative explanation of the losses scaling with the size of the square array}

The difference between the scaling of the radiative losses with the size of the square array --- $N_{tot}^{-3}$ and $N_{tot}^{-5}$ ($N_{tot} = N^2$ is the number of atoms in the array) for $A_1/B_2$ and $A_2/B_1$ states, respectively --- can be intuitively explained as follows. As we have shown in Fig.~3(d,e) in the main text, for the non-optimal periods $\tilde{a} \gtrsim 0.3$ two considered eigenstates can be approximated as $\psi^{(N,N)}$ ($A_1/B_2$ state) and $\psi^{(N-2,N)^-}$ ($A_2/B_1$ state) basis states. Since these basis states correspond to the guided waves in the corresponding infinite array, it is natural to expect that the for large size of the array the radiative losses will be determined by the intensity of the wave at the sharp edges of the array~\cite{WiersigPRL2006}, i.e. by square amplitudes of the dipole moments near the corners. For the $\psi^{(N,N)}$ state the corner dipole moment from Eq.~2 in the main text reads:
\begin{equation}
\psi^{(N,N)}_{1,1} = \dfrac{2}{\pi} q_0 \sin^2(q_0) \propto q_0^3,
\label{eq:corner_dipole_diag}
\end{equation}
where the approximation is done for small $q_0 = \pi/(N+1)$ or large $N$. Since in this case $q_0 \propto N^{-1}$, the scaling of the square amplitude of the corner dipole is $|\psi^{(N,N)}_{\text{corner}}|^2 \propto N^{-6} = N_{tot}^{-3}$ in agreement with the numerical simulations.

The corner dipole moment of the $\psi^{(N-2,N)^-}$ state is necessarily zero, since it is antisymmetric with respect to the diagonals of the square. Therefore, we consider the dipole that is next to the corner one:
\begin{equation}
\psi^{(N-2,N)^-}_{1,2} = -\dfrac{\sqrt{2}}{\pi}q_0 [\sin(q_0) \sin(6q_0) - \sin(3q_0) \sin(2q_0)] \propto q_0^5.
\label{eq:corner_dipole_odd}
\end{equation}
Due to specific symmetry of such state the first term in the expansion $\propto q_0^3$ cancels out and the scaling with $N$ becomes $|\psi^{(N-2,N)^-}_{\text{corner}}|^2 \propto N^{-10}=N_{tot}^{-5}$, again in agreement with the numerical simulations.

\section*{E. Details of scattering calculations}

We perform the scattering calculations in a semiclassical manner, where the incident wave with a given spatial distribution $\mathbf{E}_0(\mathbf{r})$ polarizes atoms, which response is described with a semiclassical atomic polarizability $\alpha(\omega) = -\frac{3 \pi}{2 k_0^2} \frac{\Gamma_0/2}{\omega-\omega_0+i\Gamma_0/2}$. As a result of re-scattering of the photons by the atoms, a dipole moment of the $j$-th atom can be found as:
\begin{equation}
\mathbf{d}^{(j)} = \mathbf{\alpha}(\omega) \mathbf{E}_0(\mathbf{r_j}) + \sum_{k \ne j} 4 \pi k_0^2 \mathbf{G}(\mathbf{R_{jk}}) \mathbf{d}^{(k)}.
\label{eq:scattering}
\end{equation}
By solving the system of linear equations Eq.~\eqref{eq:scattering} for all the atoms, we find the distribution of the dipole moments induced by the incident field, which further allows to obtain the total extinction (scattering) cross-section by using the optical theorem~\cite{Draine1988}:
\begin{equation}
\sigma_{tot} = \frac{4 \pi k_0}{ \left| \mathbf{E}_0(\mathbf{r}) \right|^2 } \sum_j \text{Im} \: \mathbf{d}^{(j)} \mathbf{E}_0^*(\mathbf{r}_j).
\label{eq:optical_theorem}
\end{equation}

To demonstrate the correspondence between spectrum of $\sigma_{tot}$ and the eigen modes $\Psi$, we decompose the dipole moment of $j$-th atom in the basis of eigen modes: $\mathbf{d}^{(j)} = \sum_{n=1}^{N_{tot}} c_n \Psi_n^{(j)}$. After substitution of this decomposition into Eq.\eqref{eq:scattering}, we obtain $\sigma_{tot} = \sum_{n=1}^{N_{tot}} \sigma_{n}$, where partial scattering cross-section $\sigma_{n}$, is defined as:
\begin{equation}
\sigma_{n} = \frac{4 \pi k_0}{ \left| \mathbf{E}_0(\mathbf{r}) \right|^2 } \sum_j \text{Im} \: c_n \Psi_n^{(j)} \mathbf{E}_0^*(\mathbf{r}_j).
\label{eq:optical_theorem_partial}
\end{equation}
Since we are considering highly non-homogeneous external fields, the denominator in Eq.~\eqref{eq:optical_theorem} strongly depends on the choice of a spatial point at which the excitation field is taken. Therefore, unlike the plane-wave excitation, the scattering cross-section in the case of the Bessel beams is defined up to a constant. To obtain some figure of merit, we normalize scattering cross-section of the system to cross-section $\sigma_0$ of a single atom positioned at maximum of longitudinal component of incident field ${E}_{0z}$.

During the calculations of scattering spectra shown at Fig.~6, we considered $\mathbf{E}_0(\mathbf{r})$ as a profile of vector Bessel beam in according with Eqs (5), (7-8) from Ref.~\cite{Kotlyar2019Oct}. We took parameters of right circular polarized beam ($s = \sigma = 1$) with orbital momenta $l = m = 7$ and $l = m = 9$, value of numerical aperture $NA = 1$, ratio of the pupil radius and the Gaussian beam waist $\beta = 0.5$.


\bibliographystyle{ieeetr}
\bibliography{ref_supp}